\documentclass{aa}
\usepackage{graphicx,amssymb}
\begin{document}
\title{A layered edge-on circumstellar disk around
HK\,Tau\,B\thanks{Based on observations carried out with the
IRAM Plateau de Bure Interferometer. IRAM is supported by
INSU/CNRS (France), MPG (Germany) and IGN (Spain).}}

   \author{G. Duch\^ene\inst{1} \and F. M\'enard\inst{2} \and
   K. Stapelfeldt\inst{3} \and G. Duvert\inst{2}}

\authorrunning{G. Duch\^ene et al.}

   \offprints{duchene@astro.ucla.edu}

   \institute{Department of Physics and Astronomy, UCLA, Los Angeles,
CA 90095-1562, USA \and
Laboratoire d'Astrophysique, Observatoire de Grenoble, Universit\'e
Joseph Fourier, BP 53, 38041 Grenoble Cedex 9, France \and
Jet Propulsion Laboratory, California Institute of Technology, 4800
Oak Grove Drive, Pasadena, CA 91109, USA}

   \date{Received 1 July 2002; accepted ....}
   
   \abstract{We present the first high angular resolution
   1.4\,mm and 2.7\,mm continuum maps of the T\,Tauri binary
   system HK\,Tau obtained with the Plateau de Bure
   Interferometer. The contributions of both components are
   well disentangled at 1.4\,mm and the star previously known
   to host an edge-on circumstellar disk, HK\,Tau\,B, is
   elongated along the disk's major axis. The optically bright
   primary dominates the thermal emission from the system at
   both wavelengths, confirming that it also has its own
   circumstellar disk. Its non-detection in scattered light
   images indicates that the two disks in this binary system
   are not parallel. Our data further indicate that the
   circumprimary disk is probably significantly smaller than
   the circumsecondary disk. \\
   We model the millimeter thermal emission from the
   circumstellar disk surrounding HK\,Tau\,B. We show that the
   disk mass derived from scattered light images cannot
   reproduce the 1.4\,mm emission using opacities of the same
   population of submicron dust grains. However, grain growth
   alone cannot match all the observed properties of this
   disk. We propose that this disk contains three separate
   layers: two thin outer surfaces which contain dust grains
   that are very similar to those of the ISM, and a disk
   interior which is relatively massive and/or has experienced
   limited grain growth with the largest grains significantly
   smaller than 1\,mm. Such a structure could naturally result
   from dust settling in a protoplanetary disk.
}

   \maketitle

\def\hk{HK\,Tau}
\def\hka{HK\,Tau\,A}
\def\hkb{HK\,Tau\,B}

%

\section{Introduction}

Circumstellar dusty disks are primarily studied to constrain
the relevant timescales and various phases of the planetary
formation process, from submicron interstellar medium (ISM)
dust material to km-sized bodies. The dramatic differences
between disks surrounding 10--20\,Myr-old zero-age main
sequence stars such as $\beta$\,Pic (Lagrange et al. 2000 and
references therein) and those around the much younger T\,Tauri
stars strongly points toward an evolutionary process that
occurs within $\lesssim10^7$\,yrs. T\,Tauri stars are thus
obvious targets for studying the very first stages of disk
evolution and, more specifically, the growth of dust grains in
protoplanetary disks from ISM-sized particules to
significantly larger ones.

The dust component of circumstellar disks can be traced
through its thermal continuum emission. This is especially
straightforward at millimeter wavelengths, at which the disk
is almost entirely optically thin. Such studies have been
performed for over a decade and first hinted at the presence
of ``large'' dust grains, as evidenced by ``abnormal'' dust
emissivity properties (Beckwith et al. 1990, hereafter
B90). However, disk masses and maximum grain size are
currently poorly constrained due to the uncertain grain shape
and composition (Pollack et al. 1994; Henning \& Stognienko
1996). Furthermore, the low spatial resolution of most radio
observations only rarely allows direct studies of the
structure of the disks nor do they resolve the tight visual
binary systems commonly found among pre-main sequence stars
(Duch\^ene 1999 and references therein). It is also possible
to study the dust grain population of circumstellar disks
through high contrast, high-angular resolution imaging of
scattered light in the environment of pre-main sequence
stars. Such observations, however, are only sensitive to those
grains whose sizes are within a factor of a few of the
wavelength. The presence (or absence) of much smaller/bigger
grains is not constrained by scattered light images. Only the
combination of the thermal and scattering approaches can fully
determine the properties of dust in circumstellar disks.

{\hk} is a 2\farcs4 pre-main sequence binary system located in
the nearby Taurus star-forming region ($D\approx140$\,pc,
Bertout et al. 1999). The primary star presents the usual
caracteristics of a classical T Tauri star surrounded by a
circumstellar accretion disk (Cohen \& Kuhi 1979, Kenyon \&
Hartmann 1995). However, the secondary component, which was
discovered by Cohen \& Kuhi (1979), is remarkably faint at
visible/near-infrared wavelengths. B90 and Beckwith et
al. (1991, hereafter BS91) first detected continuum millimeter
and submillimeter emission from the system and estimated a
first disk mass, $M_d \sim 10^{-2}\,M_\odot$, although their
large beam sizes did not resolve the binary.

The {\hk} system was later observed at high angular resolution
in the near-infrared by Koresko (1998, hereafter K98) and by
Stapelfeldt et al. (1998, hereafter S98) through deep
HST/WFPC2 visible imaging. {\hka} appeared as a bright point
source while {\hkb}\footnote{In some early studies this
component was named {\hk}/c. We adopt here the convention
proposed by the IAU and name the secondary component {\hkb}.} 
offered a dramatically different shape, that of an extended
nebulosity, about 210\,AU across, with a central dark
lane. Both interpreted these images as the signature of an
optically thick circumstellar disk seen almost edge-on that
completely obscures the central star and only lets scattered
photons reach the observer. They further estimated that the
total disk mass needed to reproduce these images assuming
ISM-like dust grains is 1--2 orders of magnitude smaller than
that derived by B90 and BS91. More recently, D'Alessio et
al. (2001, hereafter DACH01) modeled the disk of {\hkb},
trying to match simultaneously the millimeter continuum flux
of B90 and the morphology of the disk as seen in optical
scattered light. They assumed a disk mass several times larger
than that estimated by B90 and BS96 and, allowing the grain
size distribution to vary, found a reasonable match to the
data for a maximum grain size between 1\,mm and 1\,m. The
large discrepancies with the conclusions of S98 and K98 raised
doubts about the actual disk's structure and dust content.

However, to properly estimate the disk mass or to conclude
that dust grains have grown from ISM-like particules, it is
necessary to remove the ambiguity induced by the {\it
unresolved} millimeter measurement of {\hk} from B90. In this
paper, we present new 1.4\,mm and 2.7\,mm images obtained with
the IRAM Plateau de Bure Interferometer with an angular
resolution that allows us to resolve the binary system at the
shortest wavelength. The observations and results are
described in \S\,\ref{sec:obs} and \S\,\ref{sec:res}
respectively, while a model of the 1.4\,mm emission is
presented in \S\,\ref{sec:model}. In \S\,\ref{sec:discus}, we
evaluate the implications of our observations for the dust
content of the disk surrounding {\hkb} as well as for the
geometrical properties of the whole system. Finally, we
summarize our findings in \S\,\ref{sec:concl}.

%

\section{Observations and data reduction}
\label{sec:obs}

\begin{table}[t]
\caption{\label{tab:res} Flux densities and source sizes
measured in the CLEANed maps. The uncertainties quoted here
for the flux densities are those associated to the fitting
process and do not include the 10\,\% flux calibration
uncertainties; these are however included in the flux
densities quoted in the text. The deconvolved source sizes,
$a$ and $b$, are the intrinsic FWHM of each source along their
major and minor axes respectively. Upper limits are
3\,$\sigma$ for all quantities.}
\begin{center}
\begin{tabular}{cccccc}
\hline
$\lambda$ & object & $F_\nu$ & $a$ & $b$ &
PA \\ (mm) & & (mJy) & (\arcsec ) & (\arcsec ) & ($\degr$) \\
\hline
1.4 & A & 27.7$\pm$0.6 & $<0.5$ & $<0.5$ & -- \\
& B & 16.2$\pm$1.2 & 0.89$\pm$0.12 & $<0.5$ & 46$\pm$8 \\ 
\hline
2.7 & A & 6.3$\pm$0.4 & $<1$ & $<1$ & -- \\
& B & $<1.5$ & -- & -- & -- \\
\hline
\end{tabular}
\end{center}
\end{table}

\begin{figure*}[t]
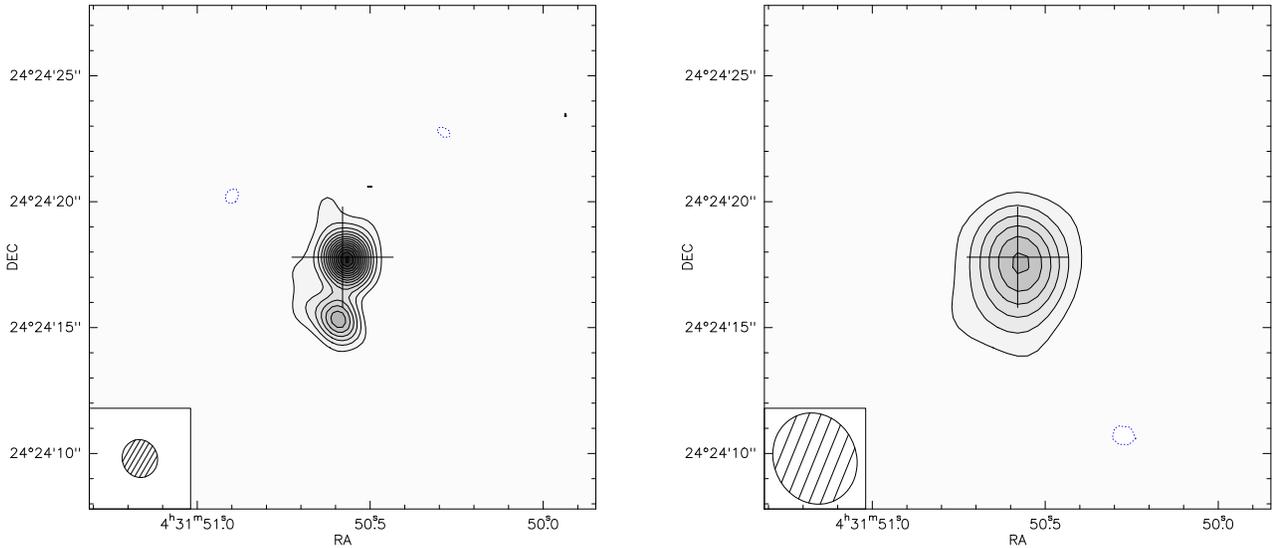

\begin{center}
  \includegraphics[width=0.4\textwidth,angle=-90]{hk-1.3-20-2.epsi}
  \hspace*{0.05\textwidth}
  \includegraphics[width=0.4\textwidth,angle=-90]{hk-2.7-20-1.epsi}
\caption{\label{fig:images} Final CLEANed maps of {\hk} at 1.4\,mm
  (left) and 2.7\,mm (right). Both images are 20{\arcsec} on a
  side. The corresponding CLEANed beam is presented in the
  small inserts. The lowest contour in both images represents
  the 2\,$\sigma$ noise level (2\,mJy/beam and 1\,mJy/beam at
  1.4\,mm and 3\,mm respectively); following contours are
  $1,2,3,\dots$ times this level. The cross indicates the
  expected location of {\hka} from Herbig \& Bell (1988),
  which is less than 0\farcs4 away from the strongest
  millimeter source.}
\end{center}
\end{figure*}

The interferometric data presented in Figure\,\ref{fig:images}
were obtained with the IRAM-Plateau de Bure interferometer
(Guilloteau et al. 1992). The array was used in compact
configurations with 4 antennas on 8 September 1998
(configuration 4D1), and 5 antennas on both 17 December 1998
(configuration 5C2), and 5 April 1999 (configuration 5D). The
baselines ranged from 24m to 176m. The measured beam size in
the final synthesized maps are 1\farcs5$\times$1\farcs4 at
position angle 15$\degr$ at 1.4\,mm (220\,GHz) and
3\farcs7$\times$3\farcs3 at position angle 23$\degr$ at
2.7\,mm (110\,GHz). Three spectral bands of 160\,MHz each at
1.4\,mm and two such spectral bands at 2.7\,mm were used. Two
more units of the correlator were tuned to observe lines of
carbon monoxyde simultaneously with the thermal continuum:
$^{13}$CO$(1\rightarrow 0)$ and $^{13}$CO$(2\rightarrow 1)$ at
110.2\,GHz (10\,MHz bandwidth) and 220.4\,GHz (20\,MHz)
respectively. None of them was detected.

We used 0415+379 and 0528+134 as phase calibrators. The rms
phase noise was 15$\degr$ to 40$\degr$ at 1.4\,mm and 8$\degr$
to 20$\degr$ at 2.7\,mm. This phase noise introduces a
position error of less than 0\farcs1. The flux scale was
calibrated by observing the sources MWC\,349, 3C\,84 and
3C\,454.4 and its associated uncertainty is about
10\,\%. Relative fluxes within a map can be obtained with a
much better accuracy. The {\sc gildas} software package was
used to reduce the data. The continuum maps were produced
using uniform weighting of the visibilities. We CLEANed the
images with a Clark algorithm using a maximum of 200 loops of
gain 0.1 each and setting the convergence level at 0.1\,\% of
the peak value.

%

\section{Results}
\label{sec:res}

Table\,\ref{tab:res} summarizes all the measurements that were
extracted from our datasets. The flux densities and positions
of the two sources were obtained through a fit to the CLEANed
images in the Fourier space. The primary, which is unresolved
at both wavelengths was fit with a point source. The secondary
was fit with a point source at 2.7\,mm and with an elliptical
gaussian at 1.4\,mm. At this shorter wavelength, {\hkb}
appears more elongated than the interferometer beam and a
point source fit systematically results in larger
residuals. From the fit to the 1.4\,mm data, we derive a
separation (2\farcs44$\pm$0\farcs05) and position angle
(171\fdg7$\pm$1\fdg1, measured East of North) between the two
centroids that are in excellent agreement with previous
astrometric results (Monetti \& Zinnecker 1991, S98). Although
the 2.7\,mm image seems to be extended roughly along the
binary orientation at the lowest contour (see
Fig\,\ref{fig:images}), the map obtained after subtraction of
the primary shows that the residuals at the location of the
secondary are not larger than several patches of noise in the
image. We therefore derived a 3\,$\sigma$ upper limit to the
actual 2.7\,mm flux density of {\hkb} from a fit to the data
in which we fixed the separation and position angle of the
binary to the values derived from the 1.4\,mm data.

The sum of the two components' fluxes at 1.4\,mm is
$43.9\pm4.6$\,mJy, in excellent agreement with B90's previous
measurement of the unresolved system
($41\pm5$\,mJy). Furthermore, we find that, although {\hka} is
the dominant source in the system, {\hkb} accounts for about
one third of the system's flux. From our data, we measure the
spectral index\footnote{This index, $\alpha_\nu$, is defined
by the relation $F_\nu\propto\nu^{\alpha_\nu}$ between 1.4 and
2.7\,mm.} of {\hka} to be $\alpha_\nu ({\rm A})=2.1\pm0.2$. To
derive a lower limit for the spectral index of {\hkb}, we
conservatively adopt a uniform probability distribution for
its 2.7\,mm flux density in the range 0--1.5\,mJy and conclude
that $\alpha_\nu ({\rm B})>3.2$ at the 99.7\,\% (3$\,\sigma$)
confidence level.

Roughly speaking, the intrinsic size of a source adds in
quadrature with the beam size to produce the observed FWHM of
the object. An increase of 0\farcs1 from the beam size can be
detected in our data. Because {\hka} is not resolved in our
data, we infer that the size of its 1.4\,mm emission is
smaller than 70\,AU (FWHM). On the contrary, {\hkb} is
extended along its major axis but unresolved along the minor
axis. We derive a size of $125\pm17$\,AU, a value smaller than
the 210\,AU disk diameter observed in the visible (S98). The
difference is likely an effect of sensitivity limit of the
interferometer to thermal dust emission. The position angle
along which the elliptical gaussian lies nicely matches the
direction of the dark lane seen in scattered light images
(40\degr, S98, K98).

%

\section{Model of the millimeter emission from {\hkb}}
\label{sec:model}

%

\subsection{Nature of the emission}

Millimeter emission from T\,Tauri stars is most frequently
thermal in nature, and it has been modelled as such in the
past (e.g., B90). However, non-thermal mechanisms, such as
free-free and gyro-synchrotron emissions, can represent a non
negligible fraction of the observed millimeter flux (e.g.,
V\,773\,Tau; Dutrey et al. 1996, hereafter D96). Since
variable non-thermal continuum emission at 6.2\,cm was
detected from {\hkb} in the past (O'Neal et al. 1990), we
discuss this possibility below.

The absence of variability in the 1.4\,mm flux density of the
unresolved system suggests that both sources are stable over
several years. This tends to indicate that non-thermal
emission, if present, is not dominant even though the
preponderant emission from {\hka} could strongly dilute the
effects of variability in {\hkb} in the single dish
measurements. Furthermore, the lower limit to the spectral
index of {\hkb} between 1.4 and 2.7\,mm is typical of
optically thin thermal emission from a circumstellar disk
(BS91, D96). Finally, the 1.4\,mm emission appears to be
elongated along the disk semi-major axis, whereas non-thermal
emission would be confined to a very small, unresolved, volume
surrounding the central star (free-free emission in the
stellar chromosphere) or elongated along an axis perpendicular
to the disk (synchrotron emission in a jet). We therefore
conclude that most of the 1.4\,mm we have detected from {\hkb}
is the result of thermal emission from dust grains in its
circumstellar disk.

%

\subsection{Model and results}
\label{subsec:model}

\subsubsection{Objectives}

To fully constrain the mass, structure and dust content of a
circumstellar disk, one ultimately needs images at all
wavelengths. Prior to our observations, the binary system had
only been resolved shortward of 3.5\,$\mu$m. All flux
measurements in the thermal regime, i.e., at longer
wavelengths, combined the flux of both stars. Building an SED
for {\hkb} is not currently possible, and previous models of
the thermal emission of this source suffered from this
limitation.

Our goal is to present an updated model of the disk around
{\hkb} that takes into account the new data of
\S\,\ref{sec:res}. Our thermal emission model is similar in
philosophy to the ones presented by B90 and D96. However, the
analytical models presented in the past assumed geometrically
thin disks and are invalid for edge-on disks. Although our
model is relatively simple, its assumptions are adequate for a
direct comparison with the scattered light model presented in
S98. Therefore, differences in the disk properties inferred
from these models, e.g., disk mass, temperature or dust
content, will indicate the limits of our understanding of this
circumstellar disk.

\subsubsection{Description of the model}

To check the relevance of the disk model derived by S98, we
adopt here the same disk structure, parametrized as follows: a
total gas+dust mass $M_d$, a surface density $\Sigma(r)\propto
r^p$ and a temperature $T(r)=T_0\,(r/r_0)^q$. The disk is
assumed to be vertically isothermal and under vertical
hydrostatic equilibrium. The dust and gaseous components of
the disk are assumed to be perfectly mixed, i.e., they have
the same temperature and density structures. Furthermore, the
dust properties are assumed to be constant throughout the disk
and a gas-to-dust mass ratio of 100:1 is assumed. We adopt a
disk outer radius of 105\,AU and an inclination of 84$\degr$
(S98). The disk inner radius is set to 0.5\,AU but has a
negligible influence on the result of our model, as the mass
enclosed inside this radius (hence the associated thermal
emission) is extremely small in all models. Finally, we
initially assume a disk temperature of $T_0=8$\,K at
$r_0=50$\,AU, which has been determined from the disk scale
height estimated by S98.

We create synthetic 1.4\,mm emission maps by integrating along
the observer's line of sight the observed flux density from
independent contiguous cells, which is determined by their
volume density, temperature, opacity and optical depth to the
observer. The resulting images are then convolved with an
elliptical gaussian kernel representing the interferometer
primary beam to compare them to our data. One of the most
crucial parameters in deriving the millimeter flux density of
a disk is the choice of the dust opacity. Following B90, we
adopt a total opacity of $\kappa_\nu (1.4\,{\rm
mm})=0.02\,{\rm cm}^2$/g which takes into account the assumed
gas-to-dust mass ratio. As noted by these authors, this value
is an order of magnitude larger than those derived for small
($a\leq0.25\,\mu$m), compact silicate grains, but seems to be
representative of the dust present in T\,Tauri circumstellar
disks. Possible explanation for this discrepancy include a
significant growth of the dust grains in circumstellar disk
(B90, BS91) or a dominant presence of submicron fractal grains
(Miyake \& Nakagawa 1993; Suttner \& Yorke 2001).

We explored a range of values for $M_d$, $p$, $q$ and $T_0$ in
an attempt to fit simultaneously the observed 1.4\,mm flux
density from {\hkb} as well as its apparent intrinsic size at
this wavelength. The latter is evaluated in our model as the
square-root difference between the FWHM of the disk and that
of the interferometer beam. We checked that this method mimics
adequately the data analysis procedures by calculating the
Fourier transform of the image of several models, sampling
them with the $uv$ coverage of the interferometer beam
corresponding to our observational configuration and fitting
the resulting datasets with the same routines as during data
reduction. The disk sizes and fluxes agreed within a few
percent in all cases.

\subsubsection{Model results}

The disk mass derived for {\hkb} by S98 ($10^{-4}\,M_\odot$)
yields a 1.4\,mm flux density of about 0.1\,mJy, two orders of
magnitude smaller than the observed one. Also, the apparent
disk size produced by this model is too small compared to our
observations. The best fit for both quantities is obtained for
a total disk mass of $M_d \approx2 \times 10^{-2}\,M_\odot$ in
our model. Larger masses result in 1.4\,mm flux densities that
are too large.

The disk mass needed to fit the thermal emission properties of
{\hkb} is excluded by the scattered light images as it would
produce a dark lane that is much too wide compared to what is
observed assuming ISM-like optical opacities and well mixed
grains. The apparent width of this dark lane basically
measures the distance between the upper and lower
$\tau_\lambda\approx1$ surfaces, and is proportional to the
product $\kappa_\lambda M_d$. Given an adopted opacity for
scattering, the total disk mass is therefore determined rather
accurately in the case of edge-on disks (e.g., Burrows et
al. 1996). We therefore conclude that the scattered light
models of S98 and the present thermal emissiom models for
{\hkb} yield solutions that are not compatible with each
other. We further discuss this discrepancy in
\S\,\ref{subsec:res_secondary}.

The disk mass derived here is similar to that derived by
B90. This was expected as the loss of a factor of three in the
observed 1.4\,mm flux density has been compensated for by the
lower disk temperature assumed in our model (8\,K vs 25\,K in
B90's model). Assuming a higher dust temperature of
$T_0=32$\,K at 50\,AU, we infer a total disk mass of
$\approx2\times10^{-3}\,M_\odot$. This value reduces the
discrepancy between scattered light and thermal emission
analyses but it is not fully satisfying. {\bf The gas scale
height for $T_0=32$\,K is twice the value derived for the dust
by S98. This could only be reconciled to the HST images by
removing the assumption that gas and dust are fully mixed
vertically, in which case the dust would have settled toward
the midplane with respect to the gas}. However, maintaining a
thermal equilibrium between the two populations is then a
difficult task, as the starlight only heats up the dust which,
in turn, keeps the gas warm through collisions.

We also find that the radial surface density and/or
temperature profiles in the disk must be relatively shallow to
match the observed ratio between the apparent size of the disk
at 1.4\,mm and its actual size ($FWHM_{1.4\,{\rm mm}}/2R_d
\approx 60$\,\%). Quantitatively, $p+q\gtrsim-1$ is required
to reproduce the apparent disk size. Models C and D from S98,
which provide the best matches to the 0.8\,$\mu$m scattered
light images, have $p+q\approx-1.1$ and $\approx-1.9$
respectively. Therefore, their model D seems too centrally
condensed to reproduce the 1.4\,mm extent of the disk.

%

\section{Discussion}
\label{sec:discus}

%

\subsection{Structure of the circumstellar disk of {\hkb}}
\label{subsec:res_secondary}

We investigate here the various disk models for {\hkb} that
could match simultaneously the scattered light and thermal
emission observations of this disk. The disk models used here
and in S98 are similar to those usually used in studies of
pre-main sequence circumstellar disks. In the past, they have
produced satisfying models of several disks (HH\,30: Burrows
et al. 1996; GG\,Tau: Guilloteau et al. 1999, McCabe et
al. 2002; IRAS\,04158+2805: M\'enard et al. 2002). However,
they fail to reproduce the observations of {\hkb}, calling for
an unusual dust model or disk structure for this source.

To reconcile the 1.4\,mm flux density of {\hkb} with the
narrow width of its dark lane in visible and near-infrared
images, the {\it ratio} of the optical-to-radio opacities must
be modified by two orders of magnitudes from the values
assumed here and in S98: either the millimeter opacity is much
larger, or the optical one is much smaller. Fractal grains are
unable to solve this dilemma, as they have larger opacities
than compact grains at both millimeter {\it and}
optical/near-infrared wavelengths (Suttner \& Yorke 2001). On
the other hand, grain growth in the disk can account for such
large changes in dust opacities: if the dust grain size
extends continuously up to mm-sized particules, the 1.4\,mm
millimeter opacity is enhanced with respect to submicron
ISM-like mixtures while the optical/near-infrared one is
greatly reduced because the mass contained in
$\sim$0.1--2\,$\mu$m grains (i.e., the most efficient
scatterers) is effectively diminished for a given disk mass. A
disk mass on the order of a few times $10^{-3}\,M_\odot$,
i.e., intermediate between the current scattered light and
thermal emission estimates, would then be favored.

There are however at least two observational constraints that
appear to rule out such a strong grain growth in this
disk. First, a grain size distribution extending up to
$a_{max}\sim1$\,mm has a millimeter spectral
index\footnote{The slightly different wavelength ranges used
in various studies to determine the spectral indices may
somewhat modify the spectral index estimates but changes are
not expected to exceed 0.1--0.2.} of $\alpha_\nu \sim2.5$--3.0
(DACH01, Wood et al. 2002), which is smaller than our
3\,$\sigma$ lower limit for {\hkb}. The observed spectral
index implies that the grain size distribution does not extend
beyond $\sim200\,\mu$m at most (Figure\,3 in DACH01). Second,
the analysis of scattered light images have allowed to
estimate an average asymmetry parameter of the dust grains,
$g_I=0.65$ (S98). This value is similar to those derived for
other pre-main sequence disks (Burrows et al. 1996, Krist et
al. 2002). Furthermore, it is very similar to the value
derived by Witt et al. (1990) for a small interstellar cloud
which presumably contains only submicron particules, whose
absorption properties match the well-known interstellar
extinction curve (Mathis \& Whiffen 1989, Kim et al. 1994,
Zubko et al. 1998). However, $g$ is a strong function of grain
size, especially for those grains with $a\sim\lambda$, with
larger grains being more forward-throwing. If there were a
significant population of such grains, uniformly distributed
throughout the disk of {\hkb}, the value of $g$ inferred from
scattered light images would therefore be significantly larger
than what is observed elsewhere. The relatively large
2.2\,$\mu$m polarization rate measured by Jensen et al. (2000)
for {\hkb} also points towards a dominant population of small
grains in the disk since large grains ($a \gg \lambda / 2\pi$)
result in largely unpolarized scattered light.

In summary, we conclude that the dust population traced by
scattered light in the disk surrounding {\hkb} does not
contain dust grains that are a few times larger
($\sim$1--3\,$\mu$m) than those found in other pre-main
sequence disks nor than in a small interstellar cloud. We
acknowledge that the presence of much bigger grains without
intermediate-size particles (i.e., two separate size
populations) cannot be established at this stage. However, it
is unclear why that intermediate-size range would be depleted:
if mm-sized grains are present and have grown from the same
population of smaller particules, then grains of any smaller
sizes should be present as well. Towards the disk midplane,
where most of the mass is, only a moderate grain growth, if
any, is allowed by the millimeter properties of the disk. {\bf
Such a grain growth is most likely unsufficient to account for
all observations of {\hkb} and would contradict our
conclusions regarding the surface of the disk, implicitely
making use of a two component (interior/surface) model.} 
Therefore, our analysis points toward an unusual disk
structure rather than to unusual dust properties.

A natural way to reconcile the scattered light images with the
1.4\,mm thermal emission map is to assume that both analyses
probe different parts of the disk that cannot be extrapolated
from one another. So far, we have assumed constant dust
properties throughout the disk as well as vertical hydrostatic
and isothermal equilibrium. This set of assumptions allowed
S98 to infer a total disk mass from the scattered light images
which, in fact, only probe the upper and lower surfaces of the
disk. If there is a sharp discontinuity in disk properties
between those surfaces and the disk interior, where most of
the mass lies and most of the thermal emission arises, the
validity of this extrapolation breaks down.

We therefore suggest that the disk surrounding {\hkb} has a
three-layer structure. The upper and lower surfaces of the
disk, above the $\tau_\lambda=1$ (as seen from the observer's
line of sight) surface at optical/near-infrared wavelength,
consist of a light layer of submicron ISM-like dust grains,
which are necessary to reproduce the scattered light
images. The central layer, which lies around the disk
midplane, has a mass in the range $\sim 10^{-3}$--$2
\times10^{-2}\,M_\odot$, depending on its temperature (but see
\S\,\ref{subsec:model}.3) and the amount of grain
growth. These various parameters cannot be distinguished on
the sole basis of a single 1.4\,mm flux measurement (e.g.,
Chiang et al. 2001).

From their fit to the optical image of {\hkb} and to the
unresolved 1.3\,mm flux measured by B90, DACH01 concluded to a
very different set of disk properties, with grain growth up to
1\,mm--1\,m. Their approach consisted in assuming a disk mass,
through an assumed mass accretion rate on the star and disk
viscosity properties, and allowing the dust properties to vary
until it matched both datasets. Their analysis is weakened by
their adopted disk mass which is arguable because {\hkb}
displays only moderate signs of accretion for instance. In any
case, the fact that this source represents only one third of
the 1.4\,mm flux density of the system would readily modify
DACH01's conclusions regarding the dust
properties. Furthermore, the spectral index of their dust
population is $\alpha_\nu \sim 2.9$--3.0, smaller than our
derived lower limit. Finally, their fit to the scattered light
image was performed with a value of $g$ fixed to the value
derived by S98, even though their grain size distribution
requires a modified value that would then be inconsistent with
the data. Overall, we believe that our observations do not
agree with their model of the disk mostly because the disk
mass and/or structure differs from their assumptions.

Models of three-layer circumstellar disks have already been
investigated by several groups (Chiang et al. 2001; DACH01;
Malbet et al. 2001) and numerical simulations of the evolution
of protoplanetary disks have also resulted in such structures
(Suttner \& Yorke 2001). Therefore, our assumption that the
disk interior differs from the surface layers is not
unphysical. Noticeably, a similar structure was also recently
suggested by Weinberger et al. (2002) for the disk surrounding
TW\,Hya on the basis of similar arguments as those presented
here. As mentionned above, however, this configuration is not
needed to match the observations of several other pre-main
sequence disks. This means that the three-layer disk structure
discussed above may not be representative of all young
circumstellar disks but may rather indicate an important
evolutionary process of circumstellar disks. Interestingly, a
similar disk structure was suggested for the more massive,
though equally young, Herbig AeBe stars by Natta et
al. (2001).

%

\subsection{The disk around {\hka}}
\label{subsec:res_primary}

Our detection of {\hka} as a millimeter thermal source
confirms that it possesses its own circumstellar disk. The
fact that we did not resolve it in our 1.4\,mm map suggests
that it is smaller than the disk surrounding {\hkb}, although
this depends on the surface density and temperature profiles
of both disks. In fact, a 100\,AU disk with power law indices
typical of other T\,Tauri circumstellar disk ($p+q \gtrsim
-2$, D96) would have an apparent size of 70--80\,AU at 2.7\,mm
(Figure\,4 in D96) and would appear even larger at 1.3\,mm
because of the larger dust emissivity and instrument
sensitivity. Therefore, the unresolved disk surrounding {\hka}
must be smaller in size than that surrounding {\hkb}, unless
it is unusually centrally concentrated. The relatively low
spectral index we derived for that source in comparison to
other T\,Tauri circumstellar disks (D96) further suggests that
the disk is partially optically thick, which would naturally
happen for a small but massive disk, although it may also
indicate that large grains are present in this disk.

The smallest disk known around a T\,Tauri star to date is the
50\,AU-radius disk around HV\,Tau\,C (Monin \& Bouvier 2000),
so this is not an unrealistically small size. However, as the
case of HV\,Tau\,C illustrates, such a disk would display an
obvious central dark lane in scattered light images if it was
inclined at $i\sim84\degr$ as {\hkb} is. To be unresolved in
the $HST$ images with such an inclination, the circumstellar
disk must be much more compact, with a radius not larger than
$\lesssim25$\,AU. Still, even though it would not be spatially
resolved, such an object would appear much fainter than it
actually is since only a small fraction of the stellar flux is
scattered towards the observer when a circumstellar disk is
seen edge-on. The observed $K$ magnitude of {\hka}
($K\approx8.5$, Kenyon \& Hartmann 1995) places it among the
majority of ``normal'' T\,Tauri stars in Taurus, while edge-on
disk sources in Taurus, such as {\hkb}, are rather
characterized by $K\gtrsim10$ (Stapelfeldt et al. 1997). We
therefore conclude that the visible and near-infrared flux
from {\hka} we receive is dominated by direct starlight as
opposed to scattered light.

In conclusion, the disk surrounding {\hka} is not seen edge-on
in contrast to the disk observed around {\hkb}. Constraining
the inclination of the disk around {\hka}, however, is not an
easy task. Our model shows that the flux from a typical disk
around a T\,Tauri star varies by less than 20\,\% when its
inclination is changed from edge-on to face-on
($i=0\degr$). Generally speaking, the observed 1.4\,mm flux
density from such a disk is insensitive to its inclination,
provided that the latter is $i\lesssim60\degr$ (see also Wood
et al. 2002). This is because the disk is mostly optically
thin at this wavelength, so that almost any part of it can be
seen from all line-of-sights. Consequently, we can only use
the fact that the star is not heavily extincted by its own
circumstellar disk and conclude that $i_{prim} \lesssim
70\degr$, which corresponds to the smallest inclination of
currently known flared edge-on disks (M\'enard et al.
2002). In other words, the two disks are not parallel to each
other and their relative inclination is $\Delta
i\gtrsim15\degr$.

In a number of simulations of the formation of binary systems
through fragmentation of an individual molecular core, a
driving role is given to rotation or magnetic field. In both
cases, the phenomenon is characterized by a prefered/symmetry
axis, leading to the formation of two stars with parallel
rotation axes, hence parallel disks. In some cases, the disks
can be parallel although they do not lie in the orbital plane
(Bonnell et al. 1992). This non-{\it coplanar} configuration
has already been suggested for {\hk} by S98 on the basis of
statistical arguments. We are now concluding that the disks
are not {\it parallel}, which breaks one more symmetry in the
system, as the stellar rotation axes are non-parallel either.
Capture of two passing-by disked T\,Tauri stars is a random
process and would naturally result in such a situation, but
this is a highly unlikely origin for binary systems in Taurus
(e.g., Clarke \& Pringle 1991). On the other hand, recent
numerical simulations show that turbulent fragmentation leads
to misaligned circumstellar disks for systems wider than
$\sim100$\,AU (M. Bate, 2002, priv. comm.), showing that this
mechanism can reasonably achieve the configuration observed in
{\hk}.

The most puzzling property of the disk surrounding {\hka}, if
confirmed, is the fact that its radius is smaller than that of
the less massive secondary component of the
system\footnote{Although pre-main sequence stellar masses are
currently strongly model-dependent, the fact that {\hkb} has a
somewhat cooler spectral type than {\hka} (Monin et al. 1998)
rules out the possibility that $M_B > M_A$.}. All theoretical
and numerical studies so far have shown that, if the disks'
size are driven by the dynamical interaction between the two
stellar components and their surrounding material, the more
massive star is always surrounded by the largest disk because
of its deeper potential well (Artymowicz \& Lubow 1994, Bate
\& Bonnell 1997). Our observations therefore suggest that the
disk sizes were established during the formation of the binary
system rather than through subsequent dynamical evolution.

%

\section{Conclusion}
\label{sec:concl}

We have obtained new high angular resolution continuum maps at
1.4\,mm and 2.7\,mm of the pre-main sequence binary system
{\hk}. For the first time, the binary is clearly resolved at
1.4\,mm and we have set an upper limit on the flux density of
the faintest component at 2.7\,mm. The emission from both
objects appears to be thermal and, while the 1.4\,mm continuum
emission from {\hkb} is resolved along the major axis of its
previsouly known edge-on circumstellar disk, {\hka} is found
to be unresolved at both wavelengths. These observations
confirm that both stars are surrounded by their own
circumstellar disk. We further find that the disk surrounding
{\hka} must be observed at a less favorable inclination
($i\lesssim70\degr$) than that of its companion, which could
be the result of turbulence-induced fragmentation, while its
radius ($\lesssim60$\,AU) is unexpectedly smaller than that of
the circumsecondary disk.

Considering the various observational constraints on the
circumstellar disk of {\hkb}, we suggest that it must have a
layered structure, such as investigated by Chiang et
al. (2001). The upper/lower layers, which are displaced
vertically from the disk midplane, do not contain a
significant population of grains larger than 1\,$\mu$m in
radius and thus have dust properties very similar to those of
the ISM. On the other hand, the interior of the disk is either
massive and constituted of ISM-like grains or has a population
of grains that extends significantly beyond $\mu$m-sized
particules. However, the grain size distribution cannot extend
to 1\,mm or beyond. Such a disk structure, although suggested
for another system, is not required for most disks around
T\,Tauri stars but could naturaly result from dust settling on
very short timescales.

Only future high-angular observations of the system covering
the mid- and far-infrared ranges will allow a complete study
of the disk structure and dust content of {\hkb}, via the
analysis of its complete SED. For instance, measurements of
the 10--20\,$\mu$m flux from {\hkb}, where the SED is close to
its minimum between the scattering and thermal regimes, would
be extremely valuable. NGST and ALMA are among the scheduled
new facilities that will allow such observations. A more
detailed study of the dust population at the surface of the
disk will also be possible through the analysis of
visible/near-infrared polarization maps of the disk.

\begin{acknowledgements}
Comments from an anonymous referee helped us significantly
improve and clarify this paper. Part of this work was
supported by the Programme National de Physique Stellaire of
the Institut National des Sciences de l'Univers. This research
has made use of the SIMBAD database, operated at CDS,
Strasbourg, France.
\end{acknowledgements}

\end{document}